\documentclass[12pt]{article}

\usepackage{graphicx}
\usepackage[margin=1.25in]{geometry}
\usepackage{amsmath}
\usepackage{amssymb}
\usepackage{amsthm}
\usepackage{bbm}
\usepackage{verbatim}
\usepackage{hyperref}
\usepackage{xcolor}
\usepackage{tikz}
\usepackage{float}
\usepackage{pgfplots}
\pgfplotsset{compat=1.18}

\usepackage[round]{natbib}   
\usepackage{advdate}

\DeclareMathOperator{\E}{\mathbb{E}}
\newcommand{\dd}{\mathrm{d}}
\newcommand{\vhat}{\hat v}

\newcommand{\one}{\mathbbm{1}}

\newtheorem{theorem}{Theorem}
\newtheorem{lemma}[theorem]{Lemma}

\newtheorem{proposition}[theorem]{Proposition}
\newtheorem{corollary}{Corollary}

\newtheorem{assumption}{Assumption}

\newtheorem{definition}{Definition}[section]
\newtheorem{example}{Example}
\newtheorem{remark}{Remark}

\title{Recommendations with Costly Verification\thanks{Gao: ying.gao@ubc.ca; Ma: ym459@cam.ac.uk. We thank Alistair Barton, Itay Fainmesser, Li Hao, and seminar and conference participants at INFORMS 2025 and the University of Cambridge for helpful comments. All errors are our own. (Draft version: October 2026. Click  \href{https://yingjupapers.s3.us-east-2.amazonaws.com/Recommendations+with+Costly+Verification.pdf}{here} for the latest version.)}}

\author{%
  Ying Gao\\{\footnotesize {University of British Columbia}} \and
  Yingju Ma\\{\footnotesize {University of Cambridge}}
}

\date{October 1, 2026}

\begin{document}

\maketitle
\begin{abstract}
A recommender faces consumers who can verify recommendations, but differ in their attention costs or in the salience of their choice of adoption. We characterize the conditions under which a technology that increases transparency by lowering the cost of verifying a product's quality will increase or decrease false recommendations. Without commitment, reducing verification costs never increases false recommendations, but when the recommender can commit to a policy, transparency can worsen both the quality of recommendations and consumer welfare. The necessary and sufficient condition is that the log-elasticity of the distribution of consumer cost types is decreasing. It can hold when the distribution is sufficiently fat-tailed, or when it mixes sophisticated and naive consumers. Our results can be applied to platform-recommended listings, security ratings, and audits.
\end{abstract}

\section{Introduction} \label{sec:introduction}
Modern markets are often organized around recommendations. Platforms and rating agencies offer rankings and suggestions of products, services and content, which affect their perception and visibility. Most of the time, these are subject to no oversight except for individual consumers' ability to fact-check them against other sources. Aside from fact-checking, the only disadvantage to the intermediary of providing false information is that they become less trustworthy. 

This paper studies the design of the content of product recommendations by a self-interested recommender when heterogeneous consumers can verify them at a cost. The recommender, e.g., a platform, can be motivated to recommend one product over another for reasons unrelated to quality, such as profit-sharing with the seller. However, frequent false recommendations make it more attractive to consumers to acquire additional information before acting. Our main finding is simple: increasing transparency to lower consumers' cost of information acquisition may or may not lead to better recommendations and increased consumer welfare, with a possible negative net effect on consumers when verification costs are heavy-tailed.

In particular, we focus on the case where the intermediary may make a binary recommendation of a preferred, risky product over a safe option (or no recommendation), which, depending on its quality, may be the better or worse option for the consumer. The intermediary knows the risky product's quality and is able to commit to their preferred recommendation policy, but cannot prevent the consumer from paying an idiosyncratic attention cost to perfectly verify the product's quality before responding to the recommendation. 

We take the canonical example of a transparency-augmenting technology to be one that lowers the cost of verification for all consumers, while fixing \emph{relative} attention-cost heterogeneity between consumers. Transparency has ambiguous effects on recommendations that depend on the distribution of consumer heterogeneity even when restricting to these cases, in which a shift in the technology affects more- and less-attentive consumers in equal proportion. We also identify the marginal effect of a general, FOSD shift in the consumer distribution: when technology improves costs in a non-proportional manner, consumers benefit most from those that decrease costs \textit{slower} for marginal consumers.

The reason is that false recommendations trade off the platform's exploitation of inframarginal adopters with persuasion of those on the margin, who may be swayed to either accept or verify recommendations depending on their level of trust in the recommender. The tradeoff depends on the log-elasticity of the consumer cost distribution because a thin population of near-accepters makes marginal improvements to the policy more appealing to the platform; decreasing the cost of verification for these near-accepters makes them into inframarginal consumers who strictly prefer to verify regardless of a change in the recommendation policy.

These effects rely on commitment. Without commitment, the recommender chooses whether to make a recommendation after observing quality, and unless it recommends every product, it makes false recommendations until the point where it is indifferent between doing so and not. The consumer who is indifferent to verifying is then fixed at the same quantile of the cost distribution, no matter the verification technology, and false recommendations become less frequent as verification gets cheaper. 

While we primarily apply the model to a platform recommending products, the framework applies wherever a better-informed, self-interested intermediary guides agents who can verify its recommendation at a heterogeneous cost. For example, e-commerce platforms promote listings that yield higher commissions, while consumers differ in the cost of acquiring outside reviews. Rating agencies are paid by the issuers whose securities they rate, while investors, from retail buyers to sophisticated institutions, have very different costs of conducting independent due diligence. Auditors certify the financial reports of the firms that hire them, while the scrutiny these statements receive from analysts and counterparties varies widely. In each case, the audience of recommender differs in the cost of verifying its recommendation. Any improvements in transparency, such as review aggregators, standardized disclosures, or open-data requirements, could lower these costs. Our results identify when such interventions discipline the recommender and when they instead lead it to make more false recommendations. The answer depends on the intermediary's commitment power and on the distribution of verification costs near the marginal verifier.

Section \ref{sec:model} describes the model of recommendations, adoption/exploration decisions, and transparency. Section \ref{sec:commitment} describes the recommendation threshold that the platform optimally commits to, and characterizes how it responds to the consumer's cost of verification, while Section \ref{sec:welfare} focuses on consumer welfare. Section~\ref{sec:nocommitment} drops commitment, shows that transparency then never increases false recommendations, and compares the two benchmarks. Finally, Section \ref{sec:conclusion} discusses non-binary recommendations and concludes. Proofs are in the Appendix.

\paragraph{Related literature}

A large literature studies information intermediaries whose incentives diverge from those of the decision makers they advise (see e.g. \citealt{Lizzeri1999}, \citealt{HagiuJullien2011}, \citealt{ArmstrongZhou2011}). Closest to us is \citet{ArmstrongChen2009}, where an exogenous share of inattentive consumers governs a seller's quality choice; we endogenize this share as the choice of consumers with different attention costs. Our concept of a recommendation lies under the framework of information design, as in \citet{KamenicaGentzkow2011}. One recent paper with a similar application is \citet{PeitzSobolev2025}, which studies a monopolist who issues personalized recommendations to consumers who differ in whether they know their own tastes, and characterizes when recommendations are biased. Unlike theirs, our recommender is uninformed about the consumer's type, and so faces heterogeneity when making a recommendation.

Attention costs have also been well-studied.
Models with informed and uninformed consumers (\citealt{SalopStiglitz1977}, \citealt{Varian1980}) share our basic tradeoff between exploiting consumers who do not check and retaining those who do, and sellers may raise verification costs deliberately (\citealt{EllisonEllison2009}, \citealt{GabaixLaibson2006}); our recommender deters verification through the content of its recommendations rather than through the cost.
\citet{Weitzman1979} and \citet{Doval2018} analyze the problem of costly search through sequential inspection, mandatory or not; our problem is simpler in that there is only one recommendation to verify, and allows us to consider the platform's design problem under a variety of cost structures. Unlike \citet{FabbriWP} and others who consider entropy-based attention costs (see \citealt{Pomattoetal}), we focus on the case of perfect verification, but instead allow for a complex cost distribution in the population.

A limited collection of other work focuses on persuasion when the receiver can endogenously obtain information. \citet{MatyskovaMontes2023} show in a more abstract persuasion environment that a sender's optimal signal is constructed to deter independent information acquisition, and that cheaper acquisition erodes the sender's persuasive power.  \citet{LukyanovSafaryan2025} study a sender's incentive to gather information to persuade an audience that fact-checks at heterogeneous costs, and finds that cheaper fact-checking coarsens the optimal public signal.

Our characterization of the optimal threshold is related to optimal monopoly pricing: the mass of false recommendations is analogous to a price, and the share of consumers who accept a recommendation without verification is analogous to the quantity sold. Our defection elasticity $\psi$ is the generalized failure rate of the cost distribution (\citealt{LariviereParteus2001}, \citealt{Lariviere2006}). It is well-known that the response of a monopolist to a multiplicative demand shift depends on the curvature of demand(\citealt{BulowPfleiderer1983}, \citealt{WeylFabinger2013}); in the language of \citet{MrazovaNeary2017}, transparency lowers standards when demand is superconvex at the optimum. Likewise, the platform's response to cost shifts depends on cost elasticity. Two features distinguish our problem. Demand is derived from consumer and product-quality heterogeneity together, so a second elasticity, the dilution elasticity $\eta$, also determines the magnitude of the platform's response. Secondly, costs have a direct effect on consumer welfare. \citet{RoeslerSzentes} elegantly describe the buyer-optimal information structure under monopoly pricing; we describe the total effect of information and costs generated by costly verification on consumer welfare.


Other work on consumer heterogeneity describes shifting and bunching effects of varying preferences. \citet{JohnsonMyatt2006} show that changes in the dispersion of consumer valuations rotate demand and that the comparative statics depend on the position of the rotation point relative to the firm's operating point; \citet{LewisSappington1994} and \citet{AndersonRenault2006} study information supplied by the seller that rotates demand. In our setting, the level and slope of the cost-growth profile at the marginal consumer play a similar role. We show that the effect of a general FOSD shift in transparency can also be broken down into these components.


\section{Model} \label{sec:model}
 
A platform chooses a product to recommend to a consumer. There are two possible products:
\begin{itemize}
    \item The \textbf{safe/standard product} ($S$) has known consumption value $v_S > 0$ to the consumer, and the platform obtains benefit $p_S$ if the consumer chooses to adopt it.
    \item The \textbf{risky product} ($R$) has a consumption value $v$ known \textit{only to the platform}. The platform obtains a benefit $p_R$ from recommending the risky product to the consumer.
\end{itemize}
 
We assume that $p_R > p_S$, that is, the platform prefers that the consumer adopt the risky product to having them adopt the safe product. There can be several possible reasons for this: the platform may want to generate reviews on a new product in order to feed its algorithm, or its share of the revenue from the risky product could be higher. It will be convenient to work with the normalized margin
\[
\gamma \equiv \frac{p_S}{p_R} \in [0,1).
\]

The consumption values $v_S$ and $v$ give a reduced-form description of the consumer's utility. We assume that $v_S \ge 0$, so that the consumer has a straightforward preference to accept the recommendation of the safe product.
 
Our focal point is the problem of the consumer when recommended the risky product: they do not know $v$, though they know the ex-ante distribution of $v$ is given by $G$; we assume it is supported on an interval of the form $[\underline{v}, \bar{v}]$. Note that $v$ can be taken as the \textit{expected} consumption value given the platform's information -- the platform need not be fully informed, only better-informed than the consumer.

 \begin{assumption}
Quality distribution $G$ is continuous, with a density $g$ that is continuous and strictly positive on $[\underline{v},0)$ and on $(0,\bar v]$, where $\underline{v}<0<\bar v$; $g$ may jump at $v=0$.
\end{assumption}

 
For products listed by a seller with an exogenous price, we could think of the consumption value as the total of the product's (possibly unknown) quality minus its (known) price, $v = q - p$. Alternatively, if the product is not priced -- e.g., it is content -- then the consumption value directly describes the consumer's welfare.
 
\paragraph{Heterogeneous verification costs}
 
The consumer is able to check the quality of a product before choosing to adopt it or not. The consumer's cost of verification depends on a privately known attention cost parameter, $x$, which is positive and distributed $\sim F_c$ in the population. Verification is perfect: the consumer learns $v$ perfectly after paying the attention cost $x$, and is then informed when choosing whether to accept or reject the risky recommendation.

We also allow the cost of verification to depend on a shared, commonly known parameter $c$ that describes the \emph{verification technology} that determines the opacity of information about the (risky) product. We assume that the single-parameter family of distributions $F_c$ describes costs that are smoothly increasing in $c$ in the sense of first-order stochastic dominance.
 

\begin{assumption}[Differentiability and FOSD]
\label{ass:F}
For every $c$, $F_c$ is an atomless distribution on $[0,\infty)$ whose support is an interval $[\underline{x}_c,\bar{x}_c]$ with $0\le\underline{x}_c<\bar{x}_c\le\infty$; the endpoints may depend on $c$, and $\underline{x}_c$ is nondecreasing in $c$. On the interior of the support, $\{(x,c):\underline{x}_c<x<\bar{x}_c\}$, $F_c(x)$ is twice continuously differentiable in $(x,c)$ with $f_c(x)>0$ and $\frac{\partial F_c(x)}{\partial c}<0$.\footnote{We allow $\underline{x}_c>0$ because natural families have this feature: a uniformly proportional family has $\underline{x}_c=c\,\varphi(0)$, and the Pareto and log-Pareto families of Section~\ref{sec:commitment} have $\underline{x}_c=c$. At the boundary $x=\underline{x}_c$ the density may jump and $F_c(x)=0$ for all $c\ge x$, so the smoothness and FOSD conditions can hold only on the interior. Remark~\ref{rem:verification-region} shows that this is innocuous: the recommender never chooses a threshold at which the posterior downside falls below $\underline{x}_c$.}
\end{assumption}
 
It will be helpful to index the consumer's type by its quantile, $q \in [0,1]$, so that the cost borne by the consumer at quantile $q$ is
\[
\kappa(c,q) = F_c^{-1}(q).
\]
 
Before proceeding further, we highlight the case where interventions in transparency (or opacity) have a uniformly proportional impact on all consumers. 
\begin{definition}
    The family $F_c$ with quantile cost function $\kappa(c,q)$ features a \emph{uniformly proportional effect of transparency} if there is some monotone $\varphi$ such that
    \begin{equation} \label{eq:proportional}
    \kappa(c,q) = c \varphi(q).
    \end{equation}
\end{definition}
The uniformly proportional case captures the benchmark in which lowering $c$ allows all consumers to learn quality $v$ more cheaply, but keeps relative heterogeneity unchanged. This assumption makes sense if consumer heterogeneity is orthogonal to the overall level of transparency -- e.g., if relative consumer heterogeneity captures heterogeneous opportunity cost of time, and $c$ captures the amount of time it takes to navigate to a third-party opinion, then technology can be represented with a family satisfying (\ref{eq:proportional}).
 
The consumer's decision proceeds as follows:
\begin{enumerate}
    \item Observe own cost parameter, $\kappa(c,q)$, and which product ($S$ or $R$) is recommended. Choose whether the initial action $a_1$ is to Accept ($A$), Reject ($R$), or Explore ($E$) the product. If $a_1 \in \{A,R\}$, the decision is over and payoffs are realized.
    \item If $a_1 = E$, pay $\kappa(c,q)$ and observe a perfect signal of $v$. Then, choose the post-exploration action $a_2 \in \{A,R\}$.
    \end{enumerate}
Denote the consumer's final decision between accept and reject (regardless of whether it is made in period $1$ or $2$) by $d$. The consumer's payoff is
\[
u_R = v \cdot \mathbbm{1}_{d = A} - \kappa(c,q) \cdot \mathbbm{1}_{a_1 = E}.
\]
Indeed, some parts of the consumer's decision are simple: upon seeing the safe product recommended (with $v_S > 0$), the consumer will choose to accept the product. However, exploration may have value when the consumer sees the risky recommendation instead.
 
We will show that their optimal decision path upon a risky recommendation depends on the platform's recommendation policy.

\paragraph{Recommendation policies}

Upon being shown the premium product, the consumer updates beliefs via Bayes Rule:
\[
\beta(v|\pi) = \frac{\pi(v) g(v)}{\int \pi(v) g(v) dv}.
\]
 
The platform's \textit{recommendation policy} is a function $\pi(v) \in [0, 1]$ that specifies the probability of recommending the premium product for each quality premium $v$.
 
The consumer's expected payoff under each initial action $a_1$, conditional on receiving a premium recommendation:
\begin{itemize}
    \item \textit{Accept}:
    \[
    U_A =\E_\beta[v].
    \]
    \item \textit{Explore}: After paying $\kappa(c,q)$ and learning $v$, the consumer accepts the premium product iff $v \ge 0$. Expected payoff:
    \[
    U_E = \E_\beta[v^+] - \kappa(c,q), \qquad \text{where } v^+ = \max(v, 0).
    \]
    \item \textit{Reject}: $U_R =0$.
\end{itemize}
 
The comparison between acceptance and rejection is governed by the posterior mean
\[
\mu(\pi)=\E_\beta[v].
\]
If $\mu(\pi)\ge0$, high-cost consumers weakly prefer accepting blindly to rejecting. In this region, the consumer population separates into explorers, those with $\kappa(c,q) < a_R(\pi)$, and blind accepters, those with $\kappa(c,q) \ge a_R(\pi)$. This is the main region studied below.
 
Exploration is valuable because it allows the consumer to avoid the negative part of the recommendation. The value of exploration relative to blind acceptance is
\[
U_E-U_A
=
\E_\beta[v^+]-\kappa(c,q)-\E_\beta[v]
=
\E_\beta[(-v)^+]-\kappa(c,q).
\]
Define the posterior downside
\[
a_R(\pi)=\E_\beta[(-v)^+].
\]
Then exploration is preferred to blind acceptance if and only if
\begin{equation}
\label{eq:exploration-cutoff}
\kappa(c,q)\le a_R(\pi).
\end{equation}
Equation~\eqref{eq:exploration-cutoff} shows that verification is selected by low-cost consumers, while high-cost consumers accept blindly, provided that blind acceptance is preferred to rejection.

\section{Optimal Recommendation under Commitment} \label{sec:commitment}
 
In the commitment benchmark, the recommender commits to a measurable policy
$\pi$ before consumers act.
This section characterizes the optimal policy and then asks how it responds to the cost of verification.
 
\subsection{Properties of the optimal policy}
A recommendation policy is a function $\pi:[\underline{v},\bar{v}]\to[0,1]$
giving the probability that the risky product is recommended at quality $v$.
Write
\[
  L(\pi) = \int \pi(v)g(v)\,dv, \qquad
  H(\pi) = \int_{v\ge 0}\pi(v)g(v)\,dv, \qquad
  B(\pi) = \int_{v < 0} \pi(v)g(v)\, dv.
\]
for the frequency of risky recommendations, the mass of good products
recommended, and the mass of bad products recommended. Denote by
\[
A(\pi) = \int_{v<0}(-v)\pi(v)g(v)\,dv
\]
the total posterior downside carried by the pool. The
posterior downside per recommendation is $a(\pi)=A(\pi)/L(\pi)$.
 
\paragraph{Acceptance/verification region.}
 
The case of commitment can be thought of as a Stackelberg game in which the platform first chooses a recommendation policy, and then the consumer chooses their response to the realized recommendation once the policy has been fixed. In every equilibrium of this Stackelberg game, the platform's recommendation induces a weakly positive posterior mean,
\[
\mu(\pi) = \frac{\int_v v \pi(v)g(v)dv}{L(\pi)} \ge 0,
\]
and a consumer rejects a recommendation with probability $0$.
 
\begin{lemma}\label{lem:accept}
    Under any optimal policy the consumer's posterior expectation of recommended quality is weakly positive, $\mu(\pi) \ge 0$, and  the platform expects the consumer to never immediately reject a recommendation.
\end{lemma}
\begin{proof}
    See Appendix.
\end{proof}
Given that the consumer either accepts or verifies the recommendation of the risky product, total demand for the risky product is
\[
D(\pi) = H(\pi) + \left(L(\pi)-H(\pi)\right)\left[1-F_c\left(\frac{A(\pi)}{L(\pi)}\right)\right].
\]
The platform's optimal choice of $\pi$ maximizes
\begin{equation} \label{eq:Pi}
\Pi(\pi) = D(\pi)p_R + (1-L(\pi))p_S.
\end{equation}
 
\paragraph{Demand under a threshold.} The platform's objective is supermodular in the pairing between recommendation probability and quality, so the optimal coupling is monotone and takes the form of a threshold quality, $\vhat$, above which the recommendation is always made and below which it is never made.
\begin{proposition}[Threshold optimality]
\label{prop:threshold}
Fix any recommendation frequency $L\in(0,1)$. Among all policies with
$\int\pi g = L$, risky demand is maximized by recommending the top $L$ mass of
qualities. Hence an optimal commitment policy may be taken to be
$\pi^*(v)=\mathbbm{1}{v\ge\vhat}$ for some threshold $\vhat$.
\end{proposition}
 
\begin{proof}
In the acceptance--verification region every recommended good product is
eventually adopted, while a recommended bad product is adopted only by
consumers who accept blindly. Hence, $D(\pi)$ is increasing in $H(\pi)$ and decreasing in $A(\pi)$ for fixed $L$.
 
Suppose $\pi$ recommends some quality $v_1$ with positive probability while
excluding some $v_2 > v_1$. Shifting recommendation probability from $v_1$ to
$v_2$ holds $L$ fixed, weakly raises $H$, and weakly lowers $A$, strictly so
whenever the shift crosses zero or removes a worse bad product. Iterating such rearrangements yields the upper-contour policy $\mathbbm{1}_{v\ge\vhat}$.
\end{proof}

For a threshold $\vhat$ write
\[
  L(\vhat) = 1-G(\vhat), \qquad
  B(\vhat) = G(0)-G(\vhat), \qquad
  H = 1-G(0),
\]
so that $H$ is the mass of good products and $B(\vhat)$ the mass of bad
products admitted into the pool. The posterior expected downside and the quantile of the marginal consumer are
\[
  a(\vhat)  =  \frac{\int_{\vhat}^{0}(-v)g(v)\,dv}{1-G(\vhat)},
  \qquad q_A(\vhat,c) = F_c(a(\vhat)),
\]
and risky demand is
\begin{equation}
  D(\vhat)  =  H + B(\vhat)\left[1-F_c(a(\vhat))\right].
  \label{eq:demand}
\end{equation}
Equation~\eqref{eq:demand} separates demand into an always-inframarginal component, the good mass $H$, which is adopted whatever the standard, and a component that is affected by the marginal consumer, the bad mass $B(\vhat)$ multiplied by the share $S(c,a)= 1-F_c(a)$ of consumers who accept blindly. Lowering the standard adds bad mass but raises the posterior downside, inducing more consumers to verify.

\begin{remark}[Verification region]
\label{rem:verification-region}
If $\underline{x}_c>0$, no consumer verifies when $a(\vhat)\le\underline{x}_c$, so $S(c,a(\vhat))=1$ and \eqref{eq:Pi} reduces to $p_R(1-G(\vhat))+p_SG(\vhat)$, which is strictly decreasing in $\vhat$ because $p_R>p_S$. The recommender therefore never stops in this region: every optimal threshold satisfies $a(\vhat^*)\ge\underline{x}_c$. Write $\vhat_c$ for the threshold with $a(\vhat_c)=\underline{x}_c$, with $\vhat_c=0$ when $\underline{x}_c=0$. The analysis below is carried out on the \emph{verification region} $\{\vhat:\vhat_0<\vhat<\vhat_c\}$, on which $a(\vhat)$ lies in the interior of the support of $F_c$, so that $S$, $\psi$ and their derivatives are well defined and smooth. At $\vhat_c$ itself $\psi$ jumps from $0$ to $\underline{x}_cf_c(\underline{x}_c^+)$, so the corner $\vhat_c$ can in principle be optimal; in all our examples the optimum is interior to the verification region, which we verify numerically.
\end{remark}
 
\subsection{An Elasticity Formula}
The recommender's problem can be seen as the following: they have access to a mass of good recommendations, which are used to subsidize the value of bad recommendations ($v < 0$). The goal is to maximize the mass of bad recommendations accepted, since the quantity of true ``good" recommendations, $H$, is fixed and always accepted.
 
For bad recommendations, there is an effective demand curve described by the tradeoff between $B(\vhat)$ and $S(c,a)$, the fraction of blind accepters. An increase in $B(\vhat)$ decreases $S(c,a)$, since verification becomes more valuable as posterior expected downside increases. Here, $B(\vhat)$ serves as a quantity margin, and $S(c,a)$ as a price, since if $S(c,a) = 0$, then each false recommendation generates $0$ profit for the recommender.
 
In contrast to standard pricing environments, the tradeoff depends not just on consumer variation, but also on the rate of dilution of the recommendation pool that corresponds to a given shift in the marginal consumer.
 
Define the \emph{defection elasticity}
\begin{equation}
  \psi(a, c)=\frac{a f_c(a)}{1-F_c(a)}=-\frac{d\ln S(c,a)}{d\ln a},
  \label{eq:psi}
\end{equation}
driven by the distribution of consumers $F$, to be the proportional rate at which consumers ``defect" to verification due to a change in the posterior downside $a$, and the \emph{dilution elasticity},
\begin{equation}
    \eta(\vhat)=\frac{d\ln B(\vhat)}{d\ln a(\vhat)}
  =\frac{a(\vhat)\bigl(1-G(\vhat)\bigr)}
  {\bigl(G(0)-G(\vhat)\bigr)\bigl(|\vhat|-a(\vhat)\bigr)},
  \label{eq:eta}
\end{equation}
driven by quality distribution $G$, to be the proportional rate of increase of the mass of bad recommendations due to the same change in the posterior downside. Their ratio is the elasticity of
the inverse demand curve,
\begin{equation}
  \xi(\vhat) = -\frac{d\ln S}{d\ln B}
   = \frac{\psi\bigl(a(\vhat), c\bigr)}{\eta(\vhat)} .
  \label{eq:eps}
\end{equation}

\begin{proposition}[Optimal recommendation threshold]
\label{prop:foc}
Any optimum in the interior of the verification region, $\vhat^*\in(\vhat_0,\vhat_c)$, satisfies

\begin{equation}
  \xi(\vhat^*) = \frac{\psi\bigl(a(\vhat^*), c\bigr)}{\eta(\vhat^*)}
   = 1-\frac{\gamma}{S(c,a^*)},
  \label{eq:foc}
\end{equation}
where $a^* = a(\vhat^*)$; the RHS is a Lerner index with price $S(c,a^*)$ and
marginal cost $\gamma$. Moreover,
\begin{equation}
  \Pi'(\vhat) = p_R g(\vhat)\,S(c,a(\vhat))\left[\xi(\vhat)-1+\frac{\gamma}{S(c,a(\vhat))}\right],
  \label{eq:Piprime}
\end{equation}
so $\Pi'$ has the sign of the bracket, and the optimum is unique whenever $\xi(\vhat)-1+\frac{\gamma}{S(c,a(\vhat))}$ is strictly decreasing in $\vhat$.
\end{proposition}

\begin{proof}
Differentiating \eqref{eq:Pi} with $G'=g$, $B'=-g$ and $S'=-f_c$,
\begin{align*}
    \begin{split}
        \Pi'(\vhat)&=p_R\left(\gamma g-gS(c,a(\vhat))+B(\vhat) f_c(a(\vhat))\bigl|a'(\vhat)\bigr|\right)\\
  &=-p_R\left(g\bigl[S(c,a(\vhat))-\gamma\bigr]-B(\vhat)f_c(a(\vhat))\bigl|a'(\vhat)\bigr|\right).
    \end{split}
\end{align*}
By \eqref{eq:eps}, $B(\vhat)f_c(a(\vhat))|a'(\vhat)|=gS(c,a(\vhat))\xi(\vhat)$, so
$\Pi'=p_Rg\bigl[S\xi-S+\gamma\bigr]$, which is \eqref{eq:Piprime}. Setting
$\Pi'=0$ and dividing by $S>0$ gives (\ref{eq:foc}).
\end{proof}

Note that since setting the optimal threshold is analogous to setting a monopoly price through false recommendations, the natural tradeoff is between marginal buyers and the deepening of the pool of bad recommendations. The term $\eta(\vhat)$ translates a shift in the marginal buyer to the impact on the mass of accepted false recommendations that the platform benefits from, so while the platform's first order condition depends on consumer heterogeneity through $\psi$ as in standard monopoly pricing, it also depends on the product quality distribution $G$ through $\eta$.
 
\begin{remark}[Regularity]
\label{rem:reg}
Uniqueness in Proposition~\ref{prop:foc} requires $\xi(\vhat)-1+\gamma/S(c,a(\vhat))$ to be strictly decreasing in $\vhat$ where it crosses zero. This is analogous to Myerson-type regularity in monopoly pricing, except that here $\psi$ is the failure rate, evaluated at the endogenous verification cutoff $a(\vhat)$.
 
The Lerner term $\gamma/S(c,a(\vhat))$ is always decreasing in $\vhat$: raising the standard lowers the posterior downside, so fewer consumers verify. But in addition, the demand elasticity $\xi(\vhat)=\psi(a(\vhat),c)/\eta(\vhat)$ must be decreasing in $\vhat$. Since $a(\vhat)$ decreases with $\vhat$, a simple sufficient condition is that (i) $\psi$ is nondecreasing in $a$, i.e. the consumer side satisfies IGFR, and (ii) the dilution elasticity $\eta$ is nondecreasing in $\vhat$. 
\end{remark}
 
\subsection{Verification Costs and Recommendation Standards}
This section discusses the effect of transparency on optimal recommendation thresholds. We think of transparency as being controlled by the verification technology $c$: when $c$ increases, verification becomes more costly for every consumer.
 
The main result shows that increased transparency need not always discipline the recommender towards fewer bad recommendations. The effect of a proportional shift in transparency when the safe option is worth $0$ to the platform (or the platform can recommend the risky product without excluding the safe one) depends on the \textit{elasticity of consumer-side elasticity $\psi$}; non-proportional shifts also affect the policy by reshaping the distribution of costs in the neighborhood of the marginal consumer.
 
To see this, note that the optimality condition can be summarized as
\begin{equation}
  \Phi(\vhat,c) = \xi(\vhat,c)-1+\frac{\gamma}{S\bigl(a(\vhat)\bigr)} = 0
  \label{eq:Phi}
\end{equation}
where we assume that we are in the case where $\Phi(\vhat,c)$ is strictly decreasing.

We can describe how unevenly a change in the technology affects the population by the \emph{cost-growth profile}
\begin{equation} \label{eq:egrowth}
e(c,q) \;\equiv\; \frac{\partial \ln \kappa(c,q)}{\partial \ln c} \;\ge\; 0,
\end{equation}
which records the elasticity with respect to $c$ of the cost borne by the consumer at quantile $q$. A family of distributions is uniformly proportional if $e(c,q) \equiv 1$, i.e., every consumer's cost rises with unit elasticity in $c$. Write
\[
  e^* = e\bigl(c,q_A\bigr), \qquad
  e'^* = \frac{\partial e(c,q)}{\partial q}\bigg|_{q=q_A}
\]
for the level and slope of the cost-growth profile \eqref{eq:egrowth} at the marginal quantile, and denote the elasticity of the defection elasticity by
\begin{equation}
  \Lambda \;\equiv\; \frac{\partial \ln \psi(a,c)}{\partial \ln a}.
  \label{eq:Lambda}
\end{equation}
Intuitively, $\Lambda$ describes the rate at which a shift in the marginal consumer $a$ thickens or thins out the density of near-marginal vs. inframarginal consumers. This affects the optimal standard because a decrease in transparency (increase in $c$) shifts the marginal consumer independently of a shift in the recommendation policy and has a first-order impact on the platform's incentive to capture vs. exploit consumers. 
 
\begin{lemma}
\label{lem:tech-partials}
At any $\vhat$,
\[
  \frac{\partial S}{\partial c}=\frac{e^*}{c}\,\psi S \;\ge\;0,
  \qquad
  \frac{\partial \psi}{\partial c}=-\frac{\psi}{c}\Bigl[e^*\Lambda+e'^*\,\psi S\Bigr].
\]
\end{lemma}

\begin{proposition}[Opacity comparative static]
\label{prop:cs}

\begin{equation}
  \operatorname{sign}(d\vhat^*/dc) = -\operatorname{sign}\left[e^*\left(\Lambda+\frac{\gamma\,\eta(\vhat^*)}{S(c,a^*)}\right)
  +e'^*\,\psi\,S(c,a^*)\right].
  \label{eq:Tstat}
\end{equation}
\end{proposition}
 
\begin{proof}
Since $\partial\Phi/\partial\vhat<0$, the implicit function theorem gives
$\operatorname{sign}(\dd\vhat^*/\dd c)=\operatorname{sign}(\partial\Phi/\partial c)$. By Lemma~\ref{lem:tech-partials},
\begin{equation} \label{eq:opacity_comp_stat}
  \frac{\partial\Phi}{\partial c}
  =\frac{1}{\eta}\frac{\partial\psi}{\partial c}-\frac{\gamma}{S^2}\frac{\partial S}{\partial c}
  =-\frac{\psi}{c\,\eta}\left\{e^*\Lambda+e'^*\psi S+\frac{\gamma e^*\eta}{S}\right\}.
\end{equation}
\end{proof}

To understand the proposition, start with the simplest case of uniformly proportional shifts with $\gamma = 0$, where direction of the policy response depends only on the sign of $\Lambda$, which could be either positive or negative. We give examples of both in the subsection to follow, and show that when $f_c$ is fat-tailed $\psi$ can be decreasing throughout the verification region, or non-monotone in $a$ with $\Lambda<0$ at $a^*$.

\begin{figure}[!t]
\centering
\caption{Opacity and the defection elasticity}
\label{fig:opacity-tails}
\vspace{4pt}
\includegraphics[width=\textwidth]{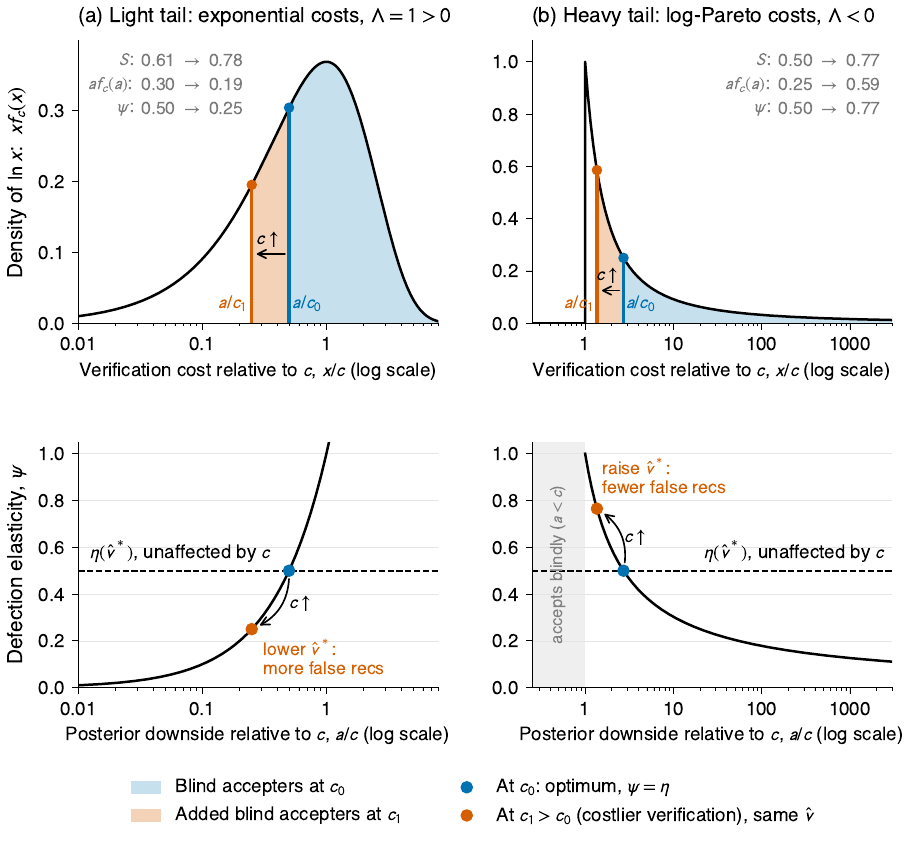}
\par\smallskip
{\footnotesize\raggedright \textit{Notes:} Proportional families, $\gamma=0$. Verification becomes costlier ($c_0\to c_1$) while the standard $\vhat$, and hence the posterior downside $a$, is held fixed. Top: density of $\ln x$; its height at the cutoff is $af_c(a)$ and the shaded area is the blind-accepter share $S$, so $\psi=af_c(a)/S$. Bottom: $\psi$ against $a/c$, with $\eta(\vhat)$ unaffected by $c$. (a) Exponential costs (Example~\ref{ex:exploit}): $\psi$ falls below $\eta$ and $\vhat^*$ falls. (b) Log-Pareto costs (Example~\ref{ex:retain}, $\nu=1$): $\psi$ rises above $\eta$ and $\vhat^*$ rises. Cutoffs are illustrative.\par}
\end{figure}

The reason is that opacity lowers the share of verifiers $q_A=F_c(a^*)$ at the current standard. When $\psi$ is increasing in $a$, this lowers $\psi$ and makes it cheaper than before for the recommender to add bad recommendations; when $\psi$ is decreasing, it raises $\psi$ and makes adding bad recommendations more expensive (Figure~\ref{fig:opacity-tails}). Assuming uniform proportionality and no opportunity cost of making a recommendation,\footnote{In particular, this model sets $\gamma = 0$ when either the safe product is unprofitable, $p_S = 0$, or any safe products can still be chosen after exploration, in which case we can normalize an effective profit $\hat p_R = p_R - p_S$ and $\hat p_S = 0$.} this shift in the marginal consumer is the only strategically relevant change for the recommender\footnote{There is no first-order effect through $\eta$, since $\eta$ is independent of $c$.}. When there is instead a positive cost of forgoing the safe recommendation ($\gamma > 0$), verifiers are particularly costly, and opacity is more prone to increase false recommendations since it decreases the number of verifiers. This is reflected in the lower, strictly negative threshold in (\ref{eq:opacity_proportional}).

\begin{corollary}[Uniformly proportional transparency]
\label{cor:cs-proportional}
If the family $F_c$ features a uniformly proportional effect of transparency, then $e^*=1$ and $e'^*=0$, so 
\begin{equation} \label{eq:opacity_proportional}
  \frac{\dd\vhat^*}{\dd c}<0
  \quad\Longleftrightarrow\quad
  \Lambda \;>\; -\,\frac{\gamma\,\eta(\vhat^*)}{S(c,a^*)} \;\le\; 0 .
\end{equation}
\end{corollary}

Outside of the proportional case, there is an additional spread effect, which describes the relative bunching or spreading of the cost distribution just left of the marginal buyer induced by the shift in the verification technology $c$. If $e'^*<0$, then costs rise faster for consumers marginally more attentive than the marginal one and the cost distribution is locally compressed around $a(\vhat)$. Equivalently, the density $f_c(a)$ rises, defection becomes more elastic, and the recommender gains a stronger incentive to persuade marginal buyers to accept the product without verification. If $e'^*>0$ the distribution is locally spread out, which means the recommender becomes able to more cheaply flood the pool with false recommendations.

This effect is fully responsible for the direction of the difference between the recommender's response to a general transparency shift vs. the comparable proportional shift if the rate of change of costs for the marginal buyer is normalized to be the same in both cases ($e^* = 1$). To see how this might be useful, take the case of comparing the uniformly proportional cost-reduction (e.g., time-savings, as explained in Sec. \ref{sec:model}) to a policy intervention that results in a uniform \textit{additive} reduction in verification costs, e.g., reducing the payment for a third-party information source, that affects the marginal buyer equally. Since ${e^*}' < 0$ for the additive intervention, it induces a change in the platform's behavior that is more beneficial to the consumer (though this does not account for the effect of the costs themselves).

 More generally, suppose that a policymaker wishes to compare the marginal effect of any two smooth transformations of the cost distribution, parametrized by families $F_c^1$ and $F_c^2$ with local cost-growth profiles $e_1^*$ and $e_2^*$ respectively.
\begin{corollary}[Comparison with proportional transparency]
\label{cor:vs-scaling}
Fix the initial cost distribution and compare a shift with profile $e_1^*$ to a shift with profile $e_2^*$. The former yields fewer false recommendations if and only if
\[
  (e^*_1-e^*_2)\left[\Lambda+\frac{\gamma\eta}{S}\right]+(e'^*_1 - e'^*_2)\,\psi S \;<\;0 .
\]
\end{corollary}

\subsubsection{Examples}

We now give a couple of example parametrizations, under the assumption of uniformly proportional families $F_c$, that illustrate the role of consumer heterogeneity in $F_c$. We also set $\gamma=0$ throughout so that the sign of the effect of a shift in opacity $c$ is the same as that of $\Lambda$.
 
\begin{example}[Light-tailed types: opacity lowers standards]
\label{ex:exploit}
Let quality be uniform, $v\sim U[-1,1]$, and let verification costs be exponential,
$F_c(x)=1-e^{-x/c}$ on $(0,\infty)$, so that $\kappa(c,q)=-c\ln(1-q)$. The defection elasticity is $\psi(a,c)=a/c$, so $\Lambda=1>0$.
The first-order condition
$\psi(a(\vhat),c)=\eta(\vhat)$ reduces to the cubic
\begin{equation}
\label{eq:exA-foc}
\vhat^2(2+|\vhat|)=2c\,(1+|\vhat|)^2 ,
\end{equation}
which has a unique root, interior if and only if $c<3/8$, with $\dd\vhat^*/\dd c<0$. Note that as $c \rightarrow 0$,  $\hat v \rightarrow 0$ as well: vanishing costs result in no false recommendations.
\end{example}
 
\begin{example}[Pareto tail: knife-edge case]
For a Pareto tail with survival $1-F_c(x)=(c/x)^\alpha$ on $x\ge c$, so that $\underline{x}_c=c$, $\psi(a,c)=\alpha$ is constant on the verification region and $\Lambda=0$. Local changes in $c$ have no first-order effect through the consumer-side elasticity.
\end{example}
 
\begin{example}[Heavy-tailed types: opacity raises standards]
\label{ex:retain}
Let bad qualities have a thin tail: on $[-1,0)$, the density is exponential in $|\vhat|$ with
scale $\tau$, $g(v)=\kappa e^{v/\tau}$, and it is uniform, $g=\tfrac12$, on $[0,1]$.\footnote{Since uniqueness comes from a joint regularity assumption on $\eta$ and $\psi$, it depends in this example on having a non-uniform quality distribution.} Let verification costs be log-Pareto, $1-F_c(x)=(1+\ln (x/c))^{-\nu}$ on $[c,\infty)$, so that $\underline{x}_c=c$ and, on the verification region $a>c$, $\psi(a,c)=\nu/(1+\ln(a/c))$ is strictly \emph{decreasing} in $a$, with $\Lambda=-1/(1+\ln(a/c))<0$; for $a<c$ nobody verifies and $\psi=0$, and Remark~\ref{rem:verification-region} shows the optimum never lies there.
When the standard is already sufficiently low, the average downside saturates at the tail mean, $a(\vhat)\to\tau/2$, while the marginal bad product keeps deteriorating, so $\psi(a,c)$ is decreasing. Note that for sufficiently low $c$, $\vhat = \underline{v}$: vanishing costs lead to maximal false recommendations.
\end{example}

 \begin{example}[Bimodal types: two groups of consumers]
\label{ex:bimodal}
Keep the quality distribution of Example~\ref{ex:retain}. Let a fraction $p$ of consumers (``sophisticated'') have exponential costs with mean $cs_1$ and the rest (``naive'') exponential costs with mean $cs_2>cs_1$, so that $1-F_c(x)=p\,e^{-x/(cs_1)}+(1-p)\,e^{-x/(cs_2)}$. The family is uniformly proportional, satisfies Assumption~\ref{ass:F}, and both components are light-tailed. Writing $\lambda_i=1/(cs_i)$ and $w(a)$ for the share of sophisticated consumers among blind accepters, the hazard rate is $h(a)=w\lambda_1+(1-w)\lambda_2$, $\psi=a\,h(a)$, and
\begin{equation}\label{eq:Lambda-bimodal}
\Lambda(a,c)=1-\frac{a\,w(1-w)(\lambda_1-\lambda_2)^2}{h(a)} .
\end{equation}
Since $w(1-w)$ is the variance of group membership among blind accepters, $\Lambda<0$ exactly where that pool is tipping from mostly sophisticated to mostly naive: a higher downside drives out the sophisticated, while those who remain are increasingly insensitive to it. With $p=0.6$, $s_1=0.05$, $s_2=1$ and $\tau=0.15$ the optimum is unique and lies in this region for $c\in(0.26,0.44)$, where $\dd\vhat^*/\dd c>0$: as $c$ falls from $0.40$ to $0.30$, $\vhat^*$ falls from $-0.230$ to $-0.465$.
\end{example}

\paragraph{Comparison.}
Both the light- and heavy-tailed examples feature interior thresholds at $c=0.05$ and $c=0.06$, so we can compare the same increase in the verification cost:
\begin{center}
\begin{tabular}{lcc|cc}
\hline
& \multicolumn{2}{c|}{Example~\ref{ex:exploit} ($\Lambda>0$)}
& \multicolumn{2}{c}{Example~\ref{ex:retain} ($\Lambda<0$)}\\
& $c=0.05$ & $c=0.06$ & $c=0.05$ & $c=0.06$\\
\hline
threshold $\vhat^*$ & $-0.266$ & $-0.296$ & $-0.727$ & $-0.590$\\
posterior downside $a^*$ & $0.028$ & $0.034$ & $0.072$ & $0.069$\\
$\Lambda$ & $+1.00$ & $+1.00$ & $-0.73$ & $-0.88$\\
false recs $B(\vhat^*)$ & $0.133$ & $0.148$ & $0.497$ & $0.491$\\
\hline
response to $c\uparrow$
& \multicolumn{2}{c|}{$\vhat^*\downarrow$/false recs $\uparrow$}
& \multicolumn{2}{c}{$\vhat^*\uparrow$/false recs $\downarrow$}\\
\hline
\end{tabular}
\end{center}
Note that the same increase in $c$ moves the two thresholds in opposite directions. An even starker illustration may be what happens as $c \downarrow 0$: in this case, the recommender has very little power to persuade the consumer, as the consumer will verify recommendations unless they are a massive cost-outlier or the recommendation is almost perfect. However, under the fat- and thin- tailed distributions the recommender is pushed to opposite extremes: in Example~\ref{ex:exploit}, with thin tails, there are few outliers and persuasion means that false recommendations disappear from the policy. In Example~\ref{ex:retain}, as $c$ vanishes, taking advantage of a fat tail becomes magnitudes more important than persuading the marginal consumer, and so the optimal policy in the limit is full exploitation.

\begin{figure}[H]
\centering
\caption{Regime switch in the two-group example}
\label{fig:two-group-switch}
\vspace{4pt}
\includegraphics[width=\textwidth]{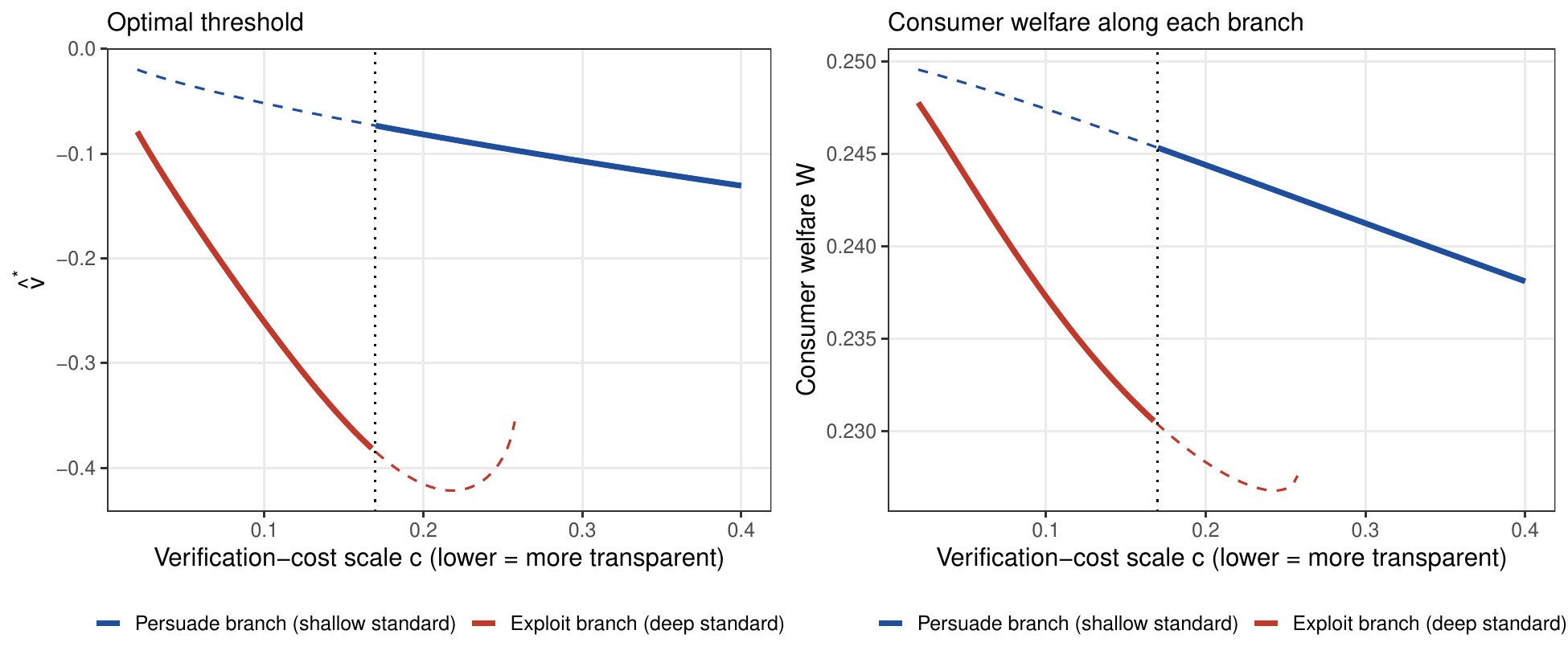}
\par\smallskip
{\footnotesize\raggedright \textit{Notes:} $p=0.7$, $s_1=0.05$, $s_2=1$, $\tau=0.15$, $\gamma=0$. Left: optimal threshold against $c$; solid segments are the global optimum, dashed segments the dominated local optimum, the dotted line the switch at $c\approx0.170$. Right: consumer welfare along each branch. Lower $c$ means cheaper verification.\par}
\end{figure}

\paragraph{Regime switches.}
With a larger sophisticated share the two-group family lowers standards discontinuously. The objective \eqref{eq:demand} splits into a persuasion term $p\,B(\vhat)\,e^{-a(\vhat)/(cs_1)}$, which peaks at a shallow standard, and an exploitation term $(1-p)\,B(\vhat)\,e^{-a(\vhat)/(cs_2)}$, which peaks at a deep one, so $\Pi$ can have two local maxima. With $p=0.7$, $s_1=0.05$, $s_2=1$ and $\tau=0.15$ the shallow branch is optimal for $c>0.170$ and the deep branch for $c<0.170$ (Figure~\ref{fig:two-group-switch}): at the switch the threshold jumps from $-0.073$ to $-0.382$, false recommendations rise from $0.193$ to $0.461$, and consumer welfare falls by $6\%$. Since $\Lambda>0$ on both branches, Proposition~\ref{prop:cs} does not directly apply to this change; the reversal comes from comparing the two regimes. Once transparency makes the sophisticated consumers verify regardless of the recommendation standard, only the naive consumers remain on the margin. Then the recommender stops maintaining a costly standard it no longer needs.

\section{Value of Transparency for Welfare} \label{sec:welfare}
For consumers who would choose to verify $v$, transparency also has a more direct positive effect: it directly lowers the disutility of acquiring information. The previous discussion showed that the behavioral effect on the platform of making $v$ more cheaply visible could, in some cases, hurt the consumer; we show here that the effect of deteriorating standards can outweigh the consumer's direct cost-savings.

Let
\[
C(\vhat,c)=\int_0^{q_A}\kappa(c,q)\,\dd q
\]
be expected verification cost per risky recommendation, integrating over the consumers who explore. Let us consider the case in which the safe product yields zero consumption utility, $v_S = 0$,\footnote{Alternatively, take $v_S=0$ as representing the case where the safe product is equivalent to the outside option.} so that consumer welfare is generated entirely by risky recommendations. Under a threshold $\vhat$ such that all consumers either accept or verify the recommendation, ex ante consumer welfare is
\begin{equation}
\label{eq:welfare-zero-safe}
W(\vhat,c)
=\int_0^{\bar v}v\,g(v)\,\dd v
\;-\;S\bigl(a(\vhat)\bigr)a(\vhat)L(\vhat)\;-\;C(\vhat,c)L(\vhat).
\end{equation}
For recommended good products, all consumers adopt the risky product, but explorers pay verification costs. For recommended bad products, blind accepters adopt and receive negative value, while explorers pay verification costs and reject.

Define the \emph{cost-weighted average growth rate among verifiers} to be
\[
\bar e\;\equiv\;\frac{\int_0^{q_A}e(c,q)\,\kappa(c,q)\,\dd q}{\int_0^{q_A}\kappa(c,q)\,\dd q}\;\ge\;0;
\]
a uniformly proportional technology has $\bar e=1$.
 
\begin{lemma}[Envelope properties]
\label{lem:envelope-general}
For any verification technology,
\[
  \frac{\partial W}{\partial\vhat}=g(\vhat)\Bigl[S\,|\vhat|+C\Bigr]>0,
  \qquad
  \frac{\partial W}{\partial c}=-L(\vhat) \frac{C\bar{e}}{c}.
\]
\end{lemma}
 
\begin{proof}
See Appendix~\ref{app:opacity-welfare}.
\end{proof}
 
Here, $\frac{\partial W}{\partial\vhat}$ captures the impact of standards on consumer welfare, while $\frac{\partial W}{\partial c}$ is the direct effect. For the net impact on consumer welfare, we combine the direct effect with the effect on $\vhat$, scaled by its impact on welfare.
\begin{proposition}[Opacity and consumer welfare]
\label{prop:opacity-welfare}
Let $v_S=0$ and let $\vhat^*(c)$ be the platform's optimal threshold. Then
\[
\frac{\dd}{\dd c}W\bigl(\vhat^*(c),c\bigr)
=\underbrace{g(\vhat^*)\Bigl[S(a^*)\,|\vhat^*|+C\Bigr]\frac{\dd\vhat^*}{\dd c}}_{\text{standards effect}}
\;-\;\underbrace{L(\vhat^*)\,\frac{C\,\bar e}{c}}_{\text{direct effect}} .
\]
\end{proposition}
 
\begin{proof}
By Lemma~\ref{lem:envelope-general} and the chain rule.
\end{proof}

Whenever standards respond positively to transparency, the two effects lie in the same direction, and transparency yields higher consumer welfare. However, when transparency makes the platform lower standards, the standards effect can dominate. In the economy of Example~\ref{ex:retain} ($\tau=0.15$, $\nu=0.3$), the following numerical example demonstrates:
\begin{center}
\begin{tabular}{lccc}
\hline
$c$ & $0.04$ & $0.05$ & $0.06$\\
\hline
$\vhat^*(c)$ & $-0.863$ & $-0.727$ & $-0.590$\\
$W(\vhat^*(c),c)$ & $0.1792$ & $0.1794$ & $0.1823$\\
$W(\vhat^*(0.04),c)$ (threshold held fixed) & $0.1792$ & $0.1780$ & $0.1769$\\
\hline
\end{tabular}
\end{center}
Consumer welfare decreases with decreasing cost $c$ at the optimum: the platform's endogenous lowering of the standard outweighs the effect of cheaper verification.\footnote{Note that $W(\vhat^*(c),c)$ is nonmonotone in $c$ here -- standards worsen still as $c \rightarrow 0$, but the direct savings effect dominates when $c$ is sufficiently low, as it does in every case.} By contrast, in the light-tailed economy of Example~\ref{ex:exploit}, $\Lambda>0$, so welfare falls monotonically in $c$.
 
\paragraph{Which Technologies Help Consumers?}
\label{sec:welfare-technology}
 
Outside the uniformly proportional case, Prop. \ref{prop:opacity-welfare} also gives some insight into which technologies are best. Clearly, if $c$ is allowed to go to $0$, consumers will obtain their first-best welfare, even if false recommendations abound. So, let us consider instead when increasing opacity (a \emph{positive} FOSD shift in $c$) is not so bad, or even helpful.
 
\begin{proposition}[Comparison.]
\label{prop:welfare-vs-scaling}
Fix the initial cost distribution. A technology with profile $e_1$ leaves consumers better off after a marginal increase in $c$ than one with profile $e_2$ does if and only if
\[
  (\bar e_1-\bar e_2)\,L\,C + \frac{\psi g(\vhat^*)\Bigl[S(a^*)|\vhat^*|+C\Bigr]}{\eta\,\bigl|\partial\Phi/\partial\vhat\bigr|}\,
  \Bigl\{(e^*_1-e^*_2)\bigl[\Lambda+\gamma\eta/S\bigr]+(e'^*_1-e'^*_2)\psi S\Bigr\}<\;0 .
\]
\end{proposition}
\begin{proof}
    See Appendix \ref{app:opacity-welfare}.
\end{proof}

For comparable shifts, we can assume that $e_1^* = e_2^*$. Note first that both $\bar e$ and ${e^*}'$ have a negative effect on consumer welfare, but there is a tension in minimizing both: when costs grow more quickly for lower-cost consumers than high-cost ones (${e^*}'$ decreases), then some verifiers must pay more ($\bar e$ must be positive, assuming costs don't decrease for anyone). A small increase in $c$ therefore benefits consumers the most if the increase in costs is sharply concentrated in near-marginal verifiers.

\section{Recommendations without Commitment} \label{sec:nocommitment}

In the commitment benchmark, the recommender chooses its policy before it observes quality. In this section, we consider the case when the recommender cannot commit to a policy. The recommender now observes $v$ first and then chooses which product to recommend. In equilibrium, consumers correctly anticipate the recommendation policy. 

\subsection{Equilibrium}

The difference from the commitment case is that the recommender no longer accounts for how its recommendations affect consumers' beliefs. Under commitment, a false recommendation raises the downside posterior and leads more consumers to verify, and the recommender weighs this cost when it chooses $\pi$. Without commitment, the recommender decides at each $v$ and takes consumers' beliefs as given. Therefore, the recommender's payoff from recommending the risky product depends on $v$ only through its sign.  A good product is adopted by all consumers, whereas a bad product is adopted only by those who do not verify, however bad it is.

Formally, a strategy for the recommender is a policy $\pi$ as in Section~\ref{sec:commitment}, where $\pi(v)$ is now the probability of recommending the risky product after observing $v$. Consumers' behavior is unchanged. They form the posterior $\beta(\cdot\mid\pi)$; when $\mu(\pi)\ge0$, a share $S(c,a(\pi))$ of them accepts without verifying. 

An \emph{equilibrium} is a policy $\pi$ such that, for every $v$, $\pi(v)$ maximizes the recommender's expected payoff given consumers' response to $\pi$.\footnote{We restrict attention to equilibria in which the risky product is recommended with positive probability. A consumer who is indifferent between accepting and rejecting accepts.}

In an equilibrium with $\mu(\pi)\ge0$, a false recommendation earns $p_R\,S(c,a(\pi))$, and the safe product earns $p_S$. The recommender is therefore indifferent between the two when the share of consumers who accept without verifying equals the normalized margin  $\gamma$. Intuitively, a false recommendation earns $p_R$ from each consumer in this share, and this must match the $p_S$ that the safe product earns. Since consumers whose cost lies below the posterior downside verify, the indifference condition holds if and only if the posterior downside equals the verification cost of the consumer at quantile $1-\gamma$. We denote this cost by
\begin{equation}\label{eq:a-gamma}
a_\gamma(c)\equiv\kappa(c,1-\gamma).
\end{equation}
It depends on $\gamma$ and $F_c$, but not on the discution $G$.

\begin{lemma}[Structure of equilibria]\label{lem:nc-structure}
In every equilibrium, consumers never reject, and every good product is recommended. Moreover, either every bad product is recommended (\emph{pooling}), or
\begin{equation}\label{eq:nc-indiff}
a(\pi)=a_\gamma(c)
\end{equation}
and the recommender earns the truthful profit $\Pi(0)=p_RH+p_SG(0)$.
\end{lemma}

The indifference condition \eqref{eq:nc-indiff} pins down the posterior downside of the pool. However, it does not pin down the composition of the pool. As a result, Lemma~\ref{lem:nc-structure} allows many equilibria.\footnote{Any pool that contains every good product, together with bad products whose average downside is $a_\gamma(c)$, is an equilibrium. One example is a pool that recommends every bad product with the same probability. How such a pool responds to $c$ depends on its composition.} We focus on threshold policies, $\pi=\one_{v\ge\vhat}$, which have the same form as the commitment policy. Note that full pooling is the threshold $\vhat=\underline v$. We write $\bar a\equiv\int_{\underline v}^0(-v)g(v)\,\dd v$ for the posterior downside under full pooling.

From now on, we assume that the safe product is profitable for the recommender, $\gamma>0$, and that the risky product is worth purchasing on average, $\E[v]\ge0$.  If $\gamma=0$, a false recommendation is never worse than the safe product, so every bad product is recommended. The second condition ensures that every threshold policy, including full pooling, has a nonnegative posterior mean, so consumers never reject it. Under these two conditions, a threshold equilibrium exists for every $c$.

\begin{proposition}[Threshold equilibrium]\label{prop:nc-eq}
There is exactly one threshold equilibrium. If $S(c,\bar a)\ge\gamma$, it is full pooling, that is, $\vhat^{nc}=\underline v$. Otherwise, its threshold $\vhat^{nc}\in(\underline v,0)$ solves $a(\vhat^{nc})=a_\gamma(c)$.
\end{proposition}

The proofs for are in Appendix~\ref{app:nocommitment}. Proposition~\ref{prop:nc-eq} says that there are two cases. Suppose that verification is so costly that many consumers accept without verifying even when every product is recommended (when  $S(c,\bar a)\ge\gamma$.).  Then every false recommendation remains profitable, so verification does not discipline the recommender, and every product is recommended. Otherwise, verification disciplines the recommender exactly up to the point of indifference.

The threshold equilibrium is also the natural one to study. Suppose that a false recommendation carries a small cost that increases with how bad the product is, for example, because worse products generate more complaints, refunds, or legal liability. Such a cost break the recommender's indifference consdition in favor of the mildest bad products and therefore eliminates other equilibria.

\begin{remark}[Equilibrium selection]\label{rem:nc-selection}
Suppose that a false recommendation of quality $v$ costs the recommender $\varepsilon h(|v|)$, where $h$ is continuous and strictly increasing, with $h(0)=0$. Then, for every $\varepsilon>0$, the equilibrium is unique and is a threshold policy. As $\varepsilon\to0$, it converges to the threshold equilibrium of Proposition~\ref{prop:nc-eq}.
\end{remark}

In every non-pooling equilibrium, the recommender earns exactly the truthful profit $\Pi(0)$, whatever $F_c$ and $c$ (Lemma~\ref{lem:nc-structure}). As long as more than a share $\gamma$ of consumers accept without verifying, a false recommendation is more profitable than the safe product. The recommender therefore adds bad products until the share falls to $\gamma$, at which point each bad product earns $p_S$ whether or not it is recommended. Blind acceptance thus generates no rent for the recommender: the profit it could earn from consumers' trust is fully dissipated.

\begin{figure}[H]
\centering
\caption{Commitment versus no commitment}
\label{fig:nc-mechanism}
\vspace{4pt}
\includegraphics[width=\textwidth]{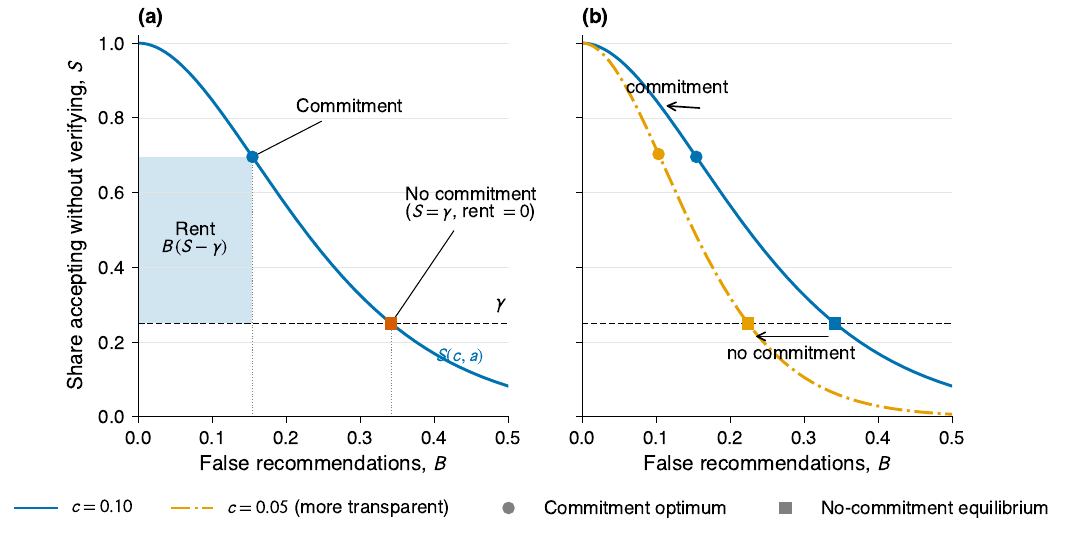}
\par\smallskip
{\footnotesize\raggedright \textit{Notes:} Exponential costs (Example~\ref{ex:exploit}) with $\gamma=1/4$ and $p_R=1$. Each curve plots the share of consumers who accept without verifying, $S(c,a(\vhat))$, against false recommendations $B(\vhat)$ as the threshold $\vhat$ varies. The recommender's profit in excess of the truthful profit equals the area $B\,(S-\gamma)$ of the rectangle under the curve and above the dashed line $S=\gamma$. Panel (a), $c=0.10$: the commitment optimum (circle) maximizes this area; the no-commitment equilibrium (square) lies at $S=\gamma$, where the area is zero. Panel (b): a decrease in $c$ from $0.10$ to $0.05$ shifts the curve down; the no-commitment equilibrium moves along $S=\gamma$ toward fewer false recommendations. Lower $c$ means cheaper verification.\par}
\end{figure}

Figure~\ref{fig:nc-mechanism} illustrates this logic. Under a threshold policy $\vhat<0$, the recommender's profit in excess of the truthful profit is
\begin{equation}\label{eq:rent}
\Pi(\vhat)-\Pi(0)=p_R\,B(\vhat)\,\bigl(S(c,a(\vhat))-\gamma\bigr),
\end{equation}
that is, $p_R$ times the area of the rectangle under the curve in panel (a) and above the line $S=\gamma$. Commitment chooses the point on the curve that maximizes this area. Without commitment, the equilibrium lies where the curve crosses $S=\gamma$, so the area is zero.

\subsection{Transparency without commitment}

Under commitment, the recommender chooses the location of the marginal verifier. Its response to transparency therefore depends on the density of consumers near that margin Without commitment, however, the marginal verifier is the consumer at quantile $1-\gamma$, whose cost rises with $c$ at the rate
\begin{equation}\label{eq:nc-threshold-cs}
\frac{\dd\ln a_\gamma(c)}{\dd\ln c}=e(c,1-\gamma)>0.
\end{equation}
The response of the equilibrium to $c$ thus depends only on how fast the cost of the consumer at quantile $1-\gamma$ grows, a rate that is positive for every technology. Neither $\Lambda$ nor $e'^*$ matters. As a result, the shape of $F_c$ affects the size of the response, but not its sign. Panel (b) of Figure~\ref{fig:nc-mechanism} illustrates the consequence: a lower $c$ shifts the curve $S(c,a(\vhat))$ down, and the equilibrium moves along the fixed line $S=\gamma$ toward fewer false recommendations.

\begin{proposition}[Transparency without commitment]\label{prop:nc-cs}
The threshold $\vhat^{nc}(c)$ and consumer welfare $W(\vhat^{nc}(c),c)$ are both nonincreasing in $c$. Moreover, both are strictly decreasing where the threshold equilibrium is not pooling.
\end{proposition}

Proposition~\ref{prop:nc-cs} shows that, without commitment, cheaper verification never increases false recommendations and never lowers consumer welfare, whatever the shape of $F_c$.\footnote{Here, $W$ is the consumer welfare of Section~\ref{sec:welfare}, where $v_S=0$. The result also holds for any $v_S\ge0$, because the additional term $G(\vhat)v_S$ is increasing in $\vhat$ and does not depend on $c$.}  In the fat-tailed case of Example~\ref{ex:retain}, the commitment optimum converges to recommending every product as $c\downarrow0$. Without commitment, however, $a_\gamma(c)=c\,\varphi(1-\gamma)\to0$ for any uniformly proportional family. As a result, the threshold equilibrium converges to recommending only good products. Therefore, the perverse effects of transparency identified in Sections~\ref{sec:commitment} also depend on the recommender's commitment power.

\subsection{The value of commitment}

Commitment is necessary for the perverse effects of transparency, but it nonetheless benefits both the recommender and consumers. To compare the two benchmarks, let $\Pi^C=\max_{\vhat}\Pi(\vhat)$ denote the commitment profit and $\Pi^{nc}=\Pi(\vhat^{nc})$ the profit in the threshold equilibrium.

\begin{proposition}[Value of commitment]\label{prop:value-commitment}
Every commitment optimum satisfies $\vhat^*\ge\vhat^{nc}$, $\Pi^C\ge\Pi^{nc}$, and $W(\vhat^*,c)\ge W(\vhat^{nc},c)$. All three inequalities are strict where the threshold equilibrium is not pooling.
\end{proposition}

The intuition follows from \eqref{eq:rent}. Commitment lets the recommender stop while the share of non-verifiers still exceeds $\gamma$. Without commitment, by contrast, that share falls to $\gamma$. As a result, the pool under commitment contains fewer bad products, so consumers also gain (Lemma~\ref{lem:envelope-general}). Commitment therefore benefits consumers in levels. However, it is also what makes their welfare respond ambiguously to transparency.

\section{Discussion and Conclusion} \label{sec:conclusion}
Before concluding, we discuss the assumption that recommendations are binary.


 


\paragraph{General recommendations}

The restriction to a binary recommendation is motivated by what is observed in practice: products tend to be shown or not shown, or alternatively, recommended or not, but rarely given a complex rating. Indeed, in our main model, we assume that the risky product can \textit{only} be accepted if it is recommended. 

Nevertheless, the question of the optimal flexible recommendation policy is interesting. In contrast to the case of persuading a single consumer with known preferences, when consumers have heterogeneous attention costs, the restriction to binary recommendations is not without loss. When recommendations can involve as many messages as necessary, the form of the optimal policy involves a single non-recommendation message and, in many cases, multiple different recommendation messages. We know that such a policy can perform strictly better than a binary one because it better couples the proportion of accepting consumers ($S$) to the proportion of recommendations under a message that are false ($B$) than what can be done with a single threshold. Formalizing such a setup is beyond our current scope, but we conjecture that the optimal policy is one in which intervals of positive and negative qualities are paired ``assortatively": recommendation messages admit a ranking in which they pool progressively shrinking intervals of more deeply negative qualities with progressively expanding intervals of milder positive qualities.

\paragraph{Conclusion}
This paper studies the content of recommendations issued by a self-interested
intermediary when the consumers who receive them can verify quality at a private
cost. The recommender is better informed, but may make false recommendations when some consumers are uninformed due to high attention costs. We have shown that an optimal binary recommendation can, perversely, respond negatively to easier verification by the consumer, via a tradeoff somewhat familiar from monopoly pricing: the consumer's cost may indeed "pass through negatively" to the recommender. Thus, starkly, recommendation quality may be worst precisely when consumers' verification capabilities are strongest. Moreover, in some regimes, the effect of deteriorating recommendations can be enough to make consumers worse-off in the presence of better verification technology.

A practical takeaway is that transparency interventions have subtle effects: at the least, estimating their effect on recommendation quality  requires knowing the shape of the attention-cost distribution near the verification cutoff, which may be nontrivial. A number of open questions remain. One, as slightly foreshadowed, is whether these results carry over to non-binary recommendation environments. Another interesting direction would be to consider the optimal design of verification technologies explicitly under practical constraints. Finally, our verification technology is stark in that it is perfectly revealing; although we reasonably believe the results carry through under other well-behaved cost and signal structures, this would be useful (and not immediate) to confirm.

\bibliography{bibliography}

\newpage
\appendix

\section{Appendix} \label{sec:app}
 
\begin{proof}[Proof of Lemma~\ref{lem:accept}]
    First, consider the case in which $\mathbb{E}[v] > 0$. In this case, $\mu(\pi) > 0$, and acceptance dominates rejection (but exploration is still possible), for any recommendation policy.

    Next, consider the case where $\mathbb{E}[v] \le 0$. Define $v_0$ to be the unique threshold such that if the platform uses the recommendation policy $\pi_0$ that recommends the product if and only if $v > \hat{v}_0$, then $\mu(\pi_0) = 0$.
 
    Now consider an arbitrary policy $\pi$. If $\mu(\pi) < 0$, then all types of the consumer either reject the product or explore it. The platform's payoff is upper-bounded by $(1-G(0))p_R + G(0) p_S$. The policy that simply recommends the product only when $v > 0$ replicates this upper bound, and under it $\mu(\pi) > 0$ so there is no rejection.
 
    Finally, suppose there is a policy $\pi$ such that $\mu(\pi)=0$ and that, conditional on receiving the risky recommendation, the consumer rejects with
probability $p>0$. We show that $\pi$ cannot be optimal.

Because $\mu(\pi)=0$,
\[
    \mathbb{E}_{\beta}[v^+]
    =\mathbb{E}_{\beta}\bigl[(-v)^+\bigr]
    =a_R(\pi).
\]
Apart from the measure-zero marginal type, consumers with
$\kappa(c,q)<a$ explore, while consumers with $\kappa(c,q)>a$ are indifferent
between accepting and rejecting. Let
\[
    s=1-F_c(a)
\]
denote the mass of consumers who do not explore. By assumption, a mass $p$ of
consumers rejects and a mass $s-p$ accepts.

We construct a strictly more profitable policy by discarding a sufficiently
small set of particularly bad false recommendations. Under the posterior
induced by $\pi$, there is a positive probability of recommended qualities
satisfying
\[
    v<0
    \qquad\text{and}\qquad
    -v>a.
\]
Indeed, the recommended pool has positive probability of $v>0$, on which
$(-v)^+=0<a$, while the posterior average of $(-v)^+$ equals $a$. Hence some
recommended negative qualities must have downside strictly greater than $a$.

Choose a measurable set $E$ of such qualities whose conditional probability under the risky recommendation is $\delta\in(0,p)$.
Let
\[
    \bar y=\mathbb{E}_{\beta}[-v\mid v\in E]>a_R(\pi),
\]
and define $\widetilde\pi$ to be identical to $\pi$, except that the risky
product is no longer recommended when $v\in E$. The original policy has total downside $L(\pi)a(\pi)$. The modified policy therefore has
posterior downside
\[
    \widetilde a
    =a_R(\widetilde\pi)
    =\frac{a(\pi)-\delta\bar y}{1-\delta}
    <a_R(\pi).
\]
Moreover, because the original posterior mean is zero and the discarded
qualities have conditional mean $-\bar y$, the posterior mean under the
modified policy is
\[
    \mu(\widetilde\pi)
    =\frac{\delta\bar y}{1-\delta}
    >0.
\]

Every consumer who rejected under $\pi$ had verification cost
$\kappa(c,q)\geq a$. Under $\widetilde\pi$, such a consumer strictly prefers
blind acceptance to rejection because $\mu(\widetilde\pi)>0$, and strictly
prefers blind acceptance to exploration because
\[
    U_E-U_A
    =\widetilde a-\kappa(c,q)
    <a_R(\pi)-\kappa(c,q)
    \leq 0.
\]
Thus, every consumer either accepts or verifies the new policy.

We have that $D(\widetilde\pi)-D(\pi)
    \geq p(1-\delta)L(\pi)-(s-p)\delta L(\pi)
    =L(\pi)(p-s\delta)$, and moreover, the platform now receives $p_S$
from the safe recommendations made on the discarded states, so
\[
    \Pi(\widetilde\pi)-\Pi(\pi)
    \geq p_R L(\pi)(p-s\delta)+p_S\delta L(\pi)
    >0.
\]
Therefore, any policy with $\mu(\pi)=0$ under which the consumer rejects with
positive probability is strictly dominated; having ruled out all the cases above, we know that any optimal policy induces rejection with probability zero.
\end{proof}

\begin{lemma}[Derivative of posterior downside]
\label{lem:downside-derivative}
For $\vhat<0$,
\[
a'(\vhat)=
-\frac{g(\vhat)}{1-G(\vhat)}\left(|\vhat|-a(\vhat)\right)<0.
\]
\end{lemma}
 
\begin{proof}
We have $a=A/L$. Since $A'(\vhat)=\vhat g(\vhat)$ and $L'(\vhat)=-g(\vhat)$,
\[
a'(\vhat)
=
\frac{A'L-AL'}{L^2}
=
\frac{g(\vhat)(\vhat L+A)}{L^2}
=
\frac{g(\vhat)(\vhat+a)}{L}
=
-\frac{g(\vhat)}{L}\left(|\vhat|-a\right).
\]
Because $a(\vhat)$ is an average downside over the pool while $|\vhat|$ is the largest downside included at the margin, $|\vhat|>a(\vhat)$ in the interior.
\end{proof}

\begin{proof}[Proof of Lem. \ref{lem:tech-partials}]
Differentiating $\kappa(c,F_c(a))=a$ in $c$ at fixed $a$ gives
$\partial_c F_c(a)=-\kappa_c/\kappa_q$, so
$\partial_c S=\kappa_c(c,q_A)f_c(a)$; since $\kappa(c,q_A)=a$ we have
$\kappa_c(c,q_A)=e^*a/c$ and $a\,f_c(a)=\psi S$, giving the first expression. For
the second, $f_c(a)=1/\kappa_q(c,q_A)$ and $\kappa_{qc}=[e'\kappa+e\kappa_q]/c$
yield
$\partial_c f_c=(f_c/c)\bigl[e^*(\psi-\Lambda)-e'^*\psi S\bigr]$, where we used
$(\kappa_{qq}/\kappa_q)\psi S=1+\psi-\Lambda$, itself obtained by
differentiating \eqref{eq:psi} in $\ln a$. Substituting into
$\partial_c\psi=(a/S)\bigl[\partial_c f_c-f_c^2\kappa_c/S\bigr]$ gives the
result.
\end{proof}

\section{Proof of Lemma~\ref{lem:envelope-general} and Proposition~\ref{prop:opacity-welfare}}
\label{app:opacity-welfare}
 
Write $W^c(\vhat,c)=\int_0^{\bar v}v\,g(v)\,\dd v-S(a)A(\vhat)-C(\vhat,c)L(\vhat)$, where $A(\vhat)=\int_{\vhat}^0(-v)g(v)\,\dd v$ is the total downside of the pool, $L(\vhat)=1-G(\vhat)$ the frequency of risky recommendations, $a=a(\vhat)=A/L$, and $q_A=F_c(a)$ so that $\kappa(c,q_A)=a$ and $C(\vhat,c)=\int_0^{q_A}\kappa(c,q)\,\dd q$.
 
\emph{Effect of the threshold.} Since $\partial_{\vhat}q_A=f_c(a)a'(\vhat)$, we have $\partial_{\vhat}S=-f_c(a)a'(\vhat)$ and $\partial_{\vhat}C=\kappa(c,q_A)\,\partial_{\vhat}q_A=a\,f_c(a)a'(\vhat)$. Differentiating in $\vhat$,
\[
\frac{\partial W^c}{\partial\vhat}=-S(a)A'(\vhat)-C\,L'(\vhat)+f_c(a)a'(\vhat)A-a\,f_c(a)a'(\vhat)L .
\]
The last two terms equal $f_c(a)\,a'(\vhat)\bigl[A-aL\bigr]=0$, because $aL=A$ by the definition of $a$. This is the envelope property of the consumer's exploration cutoff: the consumer at quantile $q_A$ is exactly indifferent between exploring and accepting blindly, so the composition shift in who verifies has no first-order welfare effect. With $A'(\vhat)=\vhat g(\vhat)$ and $L'(\vhat)=-g(\vhat)$,
\[
\frac{\partial W^c}{\partial\vhat}=g(\vhat)\Bigl[S(a)\,|\vhat|+C\Bigr]>0 .
\]
 
\emph{Effect of the cost.} Differentiating in $c$ at fixed $\vhat$, we have $\partial_c S=\kappa_c(c,q_A)f_c(a)$ and
\[
\partial_c C=\int_0^{q_A}\frac{\partial\kappa(c,q)}{\partial c}\,\dd q+\kappa(c,q_A)\,\partial_c q_A
=\mathcal K-a\,\kappa_c(c,q_A)f_c(a),
\]
using $\partial_c q_A=\partial_c F_c(a)=-\kappa_c/\kappa_q=-\kappa_c(c,q_A)f_c(a)$. The same cancellation applies: the two terms in $\kappa_c(c,q_A)f_c(a)$ cancel against $A=aL$, leaving only the direct increase in verification payments by the consumers who continue to verify,
\[
\frac{\partial W^c}{\partial c}=-L(\vhat)\,\mathcal K
=-\frac{L(\vhat)\,C\,\bar e}{c}\;\le\;0 .
\]
 
\emph{Combining.} Under the regularity condition of Remark~\ref{rem:reg}, $\vhat^*(c)$ is differentiable, and the chain rule gives the decomposition in Proposition~\ref{prop:opacity-welfare}. Substituting $\dd\vhat^*/\dd c=(\partial\Phi/\partial c)/|\partial\Phi/\partial\vhat|$, with $\partial\Phi/\partial c$ given by \eqref{eq:opacity_comp_stat}, and multiplying by $c>0$ gives the condition in Proposition~\ref{prop:welfare-vs-scaling}. If the bracket in \eqref{eq:Tstat} is nonnegative, then $\dd\vhat^*/\dd c\le0$, so both terms are nonpositive and consumer welfare weakly decreases in $c$. A welfare increase therefore requires a negative bracket together with a standards effect large enough to outweigh the direct effect. \qed

\section{Details for the Examples}
\label{app:examples}
 
\subsection{Example~\ref{ex:exploit} (light-tailed types)}
 
\emph{Quality side.} With $g=\tfrac12$, $G(0)=\tfrac12$, and $1-G(\vhat)=\tfrac{1+|\vhat|}{2}$,
\[
a(\vhat)=\frac{\int_0^{|\vhat|}t\,\tfrac12\,\dd t}{(1+|\vhat|)/2}=\frac{|\vhat|^2}{2(1+|\vhat|)},
\qquad
B(\vhat)=G(0)-G(\vhat)=\frac{|\vhat|}{2},
\qquad
\eta(\vhat)=\frac{1+|\vhat|}{2+|\vhat|},
\]
where the last expression follows from substituting $a(\vhat)$ into the definition of $\eta$. Note $\mu(\vhat)=\tfrac{1-|\vhat|}{2}\ge0$ for all $|\vhat|\le1$, so the entire range lies in the acceptance--verification region and $\vhat_0=\underline{v}=-1$.
 
\emph{Consumer side.} With $F_c(x)=1-e^{-x/c}$ we have $\kappa(c,q)=-c\ln(1-q)$, so the family is uniformly proportional, and $\psi(a,c)=a/c$, strictly increasing in $a$, with $\Lambda=1$.
 
\emph{Uniqueness and interiority.} The ratio $\sigma(\vhat)=\psi(a(\vhat),c)/\eta(\vhat)=\tfrac{|\vhat|^2(2+|\vhat|)}{2c(1+|\vhat|)^2}$ satisfies
\[
\frac{\dd\ln \sigma}{\dd|\vhat|}
=\frac{2}{|\vhat|}+\frac{1}{2+|\vhat|}-\frac{2}{1+|\vhat|}
=\frac{4+3|\vhat|+|\vhat|^2}{|\vhat|(1+|\vhat|)(2+|\vhat|)}>0,
\]
so \eqref{eq:exA-foc} has a unique root and the single-crossing condition of Remark~\ref{rem:reg} holds. The root is interior ($|\vhat^*|<1$) if and only if $\sigma(-1)>1$, that is, $c<3/8$. Implicit differentiation of \eqref{eq:exA-foc} gives
\[
\frac{\dd|\vhat^*|}{\dd c}
=\frac{2(1+|\vhat^*|)^2}{(4|\vhat^*|+3|\vhat^*|^{2})-4c(1+|\vhat^*|)}>0,
\]
the denominator being positive by the same single-crossing property.
 
\emph{Numerical solutions.}
\begin{center}
\begin{tabular}{lcccc}
\hline
$c$ & $0.05$ & $0.10$ & $0.15$ & $0.20$\\
\hline
$\vhat^*$ & $-0.266$ & $-0.405$ & $-0.526$ & $-0.638$\\
$a^*$ & $0.028$ & $0.058$ & $0.091$ & $0.124$\\
false recs $B(\vhat^*)$ & $0.133$ & $0.203$ & $0.263$ & $0.319$\\
demand $D(\vhat^*)$ & $0.576$ & $0.613$ & $0.644$ & $0.671$\\
\hline
\end{tabular}
\end{center}
Each optimum is verified to dominate the corner policies $\vhat=0$ (demand $\tfrac12$) and $\vhat=-1$.
 
\subsection{Example~\ref{ex:retain} (heavy-tailed types)}
 
The bad-side density is $g(v)=\kappa e^{v/\tau}$ on $[-1,0)$ with $\kappa=[2\tau(1-e^{-1/\tau})]^{-1}$, so bad products have total mass $\tfrac12$ concentrated within depth $\tau$ of zero; the good side is uniform, $g=\tfrac12$ on $[0,1]$. Verification costs satisfy $x\ge c$, so on the region $a(\vhat)<c$ nobody verifies and demand is simply $1-G(\vhat)$. The optima reported below are verified to lie in the interior verification region $a(\vhat)>c$.
 
\emph{Quality side.} Writing $\chi(\vhat)=e^{-|\vhat|/\tau}$ and $N=1-e^{-1/\tau}$,
\[
B(\vhat)=\frac{1-\chi(\vhat)}{2N},
\qquad
a(\vhat)=\frac{\tau\left[1-\left(1+\tfrac{|\vhat|}{\tau}\right)\chi(\vhat)\right]}{N\left[1+2B(\vhat)\right]},
\qquad
\eta(\vhat)=\frac{\tau\left[1-\left(1+\tfrac{|\vhat|}{\tau}\right)\chi(\vhat)\right]}{2N\,B(\vhat)\,\bigl(|\vhat|-a(\vhat)\bigr)} .
\]
For $|\vhat|$ beyond a few multiples of $\tau$ the exponential terms vanish: $a(\vhat)\to\tau/2$ and $\eta(\vhat)\approx\tau/(|\vhat|-\tau/2)$, strictly decreasing in $|\vhat|$. The economics: the average downside saturates at the tail mean $\tau/2$, while the marginal bad product at depth $|\vhat|$ keeps deteriorating, so the pool's tolerance for dilution falls as the standard deepens. The posterior mean satisfies $\mu(\vhat)\ge0.17$ for all $|\vhat|\le1$, so the acceptance--verification region is again the entire range.
 
\emph{Uniqueness.} On the verification region the first-order condition is
\[
\frac{\nu}{1+\ln\bigl(a(\vhat)/c\bigr)}=\eta(\vhat).
\]
Because $a(\vhat)$ saturates, the left side is nearly flat in $|\vhat|$ (its slope is exponentially small in $|\vhat|/\tau$), while the right side is strictly decreasing; hence $\Delta(\vhat)=\psi-\eta$ is strictly increasing where it crosses zero, which is the second-order condition for a maximum, and the crossing is unique. Substituting the saturated values gives the approximate closed form reported in the text.
 
\emph{Numerical solutions} ($\tau=0.15$, $\nu=0.3$):
\begin{center}
\begin{tabular}{lccc}
\hline
$c$ & $0.04$ & $0.05$ & $0.06$\\
\hline
$\vhat^*$ & $-0.864$ & $-0.727$ & $-0.590$\\
$a^*$ & $0.074$ & $0.072$ & $0.069$\\
false recs $B(\vhat^*)$ & $0.499$ & $0.497$ & $0.491$\\
demand $D(\vhat^*)$ & $0.933$ & $0.953$ & $0.973$\\
\hline
\end{tabular}
\end{center}
Each $a^*>c$, so $\Lambda=-1/(1+\ln(a^*/c))<0$ at every optimum, and each solution is verified to be the global maximum (against the corners and against a fine grid over $\vhat$). The false-recommendation mass falls only slightly because the bad tail is thin beyond depth $\tau$; the economically meaningful movement is in the marginal quality admitted.

\section{Proofs for Section~\ref{sec:nocommitment}}
\label{app:nocommitment}

\begin{proof}[Proof of Lemma~\ref{lem:nc-structure}]
Fix an equilibrium $\pi$ with $L(\pi)>0$.

\emph{Step 1: $\mu(\pi)\ge0$.} Suppose instead that $\mu(\pi)<0$. Then consumers who do not verify reject the risky recommendation. Hence a recommended bad product is adopted by nobody, and recommending it yields $0<p_S$. Therefore, no bad product is recommended, which implies that $\mu(\pi)>0$, a contradiction. Since $\mu(\pi)\ge0$, no consumer rejects.

\emph{Step 2: good products.} Every consumer adopts a recommended good product. Verifiers adopt it because $v\ge0$, and the others adopt it blindly. Recommending a good product therefore yields $p_R>p_S$, so $\pi(v)=1$ for all $v\ge0$.

\emph{Step 3: bad products.} Recommending a bad product yields $p_R S(c,a(\pi))$, which does not depend on $v$. If $S(c,a(\pi))>\gamma$, then $\pi(v)=1$ for almost every $v<0$, so $\pi\equiv1$. If $S(c,a(\pi))<\gamma$, then $\pi(v)=0$ for almost every $v<0$. In that case, $a(\pi)=0$. However, $S(c,0)=1>\gamma$ because $F_c$ is atomless with $F_c(0)=0$, a contradiction. Otherwise, $S(c,a(\pi))=\gamma$, which is \eqref{eq:nc-indiff}.

\emph{Step 4: profit.} In a non-pooling equilibrium, the recommender is indifferent at every bad $v$. Hence each bad state yields $p_S$ in expectation, whatever the realized recommendation. Each good state yields $p_R$. Summing over states gives $\Pi(0)=p_RH+p_SG(0)$.
\end{proof}

\begin{proof}[Proof of Proposition~\ref{prop:nc-eq}]
Under $\pi\equiv1$, the posterior is the prior, so $\mu=\E[v]\ge0$ and $a=\bar a$. The recommender does not deviate at a bad $v$ if and only if $p_RS(c,\bar a)\ge p_S$, that is, $S(c,\bar a)\ge\gamma$.

Next, consider a threshold $\vhat\in(\underline v,0]$. It truncates the prior from below, so $\mu(\vhat)\ge\E[v]\ge0$. By Lemma~\ref{lem:nc-structure}, it is an equilibrium if and only if $a(\vhat)=a_\gamma(c)$. Indeed, indifference makes $\one_{v\ge\vhat}$ optimal at every $v$. By Lemma~\ref{lem:downside-derivative}, $a(\cdot)$ is continuous and strictly decreasing on $[\underline v,0]$, from $\bar a$ to $0$. Since $a_\gamma(c)>0$, a solution in $(\underline v,0)$ therefore exists if and only if $a_\gamma(c)<\bar a$. When it exists, it is unique.

Finally, $a_\gamma(c)<\bar a$ if and only if $S(c,\bar a)<\gamma$. This is because $S(c,\cdot)$ is nonincreasing and $S(c,a_\gamma(c))=\gamma$. Moreover, $S(c,\cdot)$ is strictly decreasing to the right of $a_\gamma(c)$, which lies in the interior of the support. Hence exactly one of the two threshold equilibria exists.
\end{proof}

\begin{proof}[Proof of Remark~\ref{rem:nc-selection}]
With the cost $\varepsilon h(|v|)$, recommending the risky product at $v<0$ yields $p_RS(c,a(\pi))-\varepsilon h(|v|)$. This payoff is strictly decreasing in $|v|$. Good products still yield $p_R>p_S$, and Step 1 of the proof of Lemma~\ref{lem:nc-structure} applies unchanged. Hence every equilibrium is, up to a null set, a threshold policy. Define
\[
D_\varepsilon(\vhat)\equiv S(c,a(\vhat))-\gamma-\frac{\varepsilon}{p_R}\,h(|\vhat|).
\]
A threshold $\vhat\in(\underline v,0)$ is an equilibrium if and only if $D_\varepsilon(\vhat)=0$. The threshold $\underline v$ is an equilibrium if and only if $D_\varepsilon(\underline v)\ge0$. Note that $D_\varepsilon$ is continuous and strictly increasing in $\vhat$, because $S(c,a(\vhat))$ is nondecreasing and $h(|\vhat|)$ is strictly decreasing in $\vhat$. Moreover, $D_\varepsilon(\vhat)\to1-\gamma>0$ as $\vhat\uparrow0$. Therefore, there is exactly one equilibrium. It is $\underline v$ if $D_\varepsilon(\underline v)\ge0$, and the unique zero of $D_\varepsilon$ otherwise.

As $\varepsilon\to0$, $D_\varepsilon\to D_0$ uniformly on $[\underline v,0]$. Moreover, $D_0$ is strictly increasing near its zero $\vhat^{nc}$, because $a_\gamma(c)$ lies in the interior of the support. Hence the equilibrium threshold converges to $\vhat^{nc}$. If $D_0(\underline v)>0$, the equilibrium is full pooling for all small $\varepsilon$.
\end{proof}

\begin{proof}[Proof of Proposition~\ref{prop:nc-cs}]
First, suppose that the threshold equilibrium is not pooling. Then $a(\vhat^{nc})=a_\gamma(c)=\kappa(c,1-\gamma)$, and \eqref{eq:egrowth} gives \eqref{eq:nc-threshold-cs}. This elasticity is strictly positive because $\partial F_c/\partial c<0$ on the interior of the support. Since $a(\cdot)$ is strictly decreasing (Lemma~\ref{lem:downside-derivative}), it follows that
\[
\frac{\dd\vhat^{nc}}{\dd c}=-\frac{e(c,1-\gamma)\,a_\gamma(c)\,L(\vhat^{nc})}{c\,g(\vhat^{nc})\bigl(|\vhat^{nc}|-a_\gamma(c)\bigr)}<0.
\]
Next, $S(c,\bar a)$ is nondecreasing in $c$ by Assumption~\ref{ass:F}. Hence the values of $c$ with $S(c,\bar a)\ge\gamma$ form an upper interval, on which the equilibrium is pooling and $\vhat^{nc}=\underline v$. At the lower end of this interval, $a_\gamma(c)=\bar a$, so $\vhat^{nc}$ is continuous in $c$. Therefore, $\vhat^{nc}$ is nonincreasing everywhere.

For welfare, the chain rule gives $\frac{\dd}{\dd c}W(\vhat^{nc}(c),c)=\partial_{\vhat}W\cdot\dd\vhat^{nc}/\dd c+\partial_cW$. By Lemma~\ref{lem:envelope-general}, $\partial_{\vhat}W>0$ and $\partial_cW\le0$. Both terms are therefore nonpositive. Moreover, the first term is strictly negative where the equilibrium is not pooling.
\end{proof}

\begin{proof}[Proof of Proposition~\ref{prop:value-commitment}]
Suppose first that the threshold equilibrium is not pooling, so that $\Pi^{nc}=\Pi(0)$ by Lemma~\ref{lem:nc-structure}. As $\vhat\uparrow0$, $a(\vhat)\to0$ and $S(c,a(\vhat))\to1>\gamma$. Moreover, $B(\vhat)>0$ for every $\vhat<0$. By \eqref{eq:rent}, it follows that $\Pi(\vhat)>\Pi(0)$ for $\vhat$ close to zero, so $\Pi^C>\Pi^{nc}$. Since $\Pi(\vhat^*)>\Pi(0)$, \eqref{eq:rent} implies that $B(\vhat^*)\bigl[S(c,a^*)-\gamma\bigr]>0$, so that $S(c,a^*)>\gamma=S(c,a_\gamma(c))$. Because $S(c,\cdot)$ is nonincreasing, $a^*<a_\gamma(c)=a(\vhat^{nc})$. Because $a(\cdot)$ is strictly decreasing, $\vhat^*>\vhat^{nc}$.

Suppose instead that the threshold equilibrium is pooling. Then $\vhat^{nc}=\underline v\le\vhat^*$. Moreover, $\Pi^C\ge\Pi(\underline v)=\Pi^{nc}$, because full pooling is feasible under commitment.

In both cases, $W(\vhat^*,c)\ge W(\vhat^{nc},c)$, because $W(\cdot,c)$ is strictly increasing (Lemma~\ref{lem:envelope-general}). The inequality is strict when $\vhat^*>\vhat^{nc}$.
\end{proof}

\end{document}